# Recent Developments on Colloidal Deposits Obtained by Evaporation of Sessile Droplets on a Solid Surface

Nagesh D. Patil[a,*], Rajneesh Bhardwaj[b,*],

[a]Department of Mechanical Engineering, University of British Columbia, Vancouver, V6T 1Z4, BC, Canada.
[b]Department of Mechanical Engineering, Indian Institute of Technology Bombay, Mumbai, 400076 India.
[*]Corresponding author (email: nagesh.patil@ubc.ca, rajneesh.bhardwaj@iitb.ac.in)

*Abstract*

Understanding flow patterns and coupled transport phenomena during evaporation of droplets loaded with colloidal particles is central to design technical applications such as organizing proteins/DNA on a solid surface. We review recent reports on evaporating sessile droplets of colloidal suspensions on a solid surface. Starting from the classical mechanism of formation of a ring-like deposit, we discuss the influence of several problem parameters. Notably, thermal or solutal Marangoni effect, particle size, particle concentration, particle shape, substrate wettability, pH of the suspension etc have been found important in controlling the deposition pattern. The deposit pattern complexity and shape have been attributed to the underlying coupled transport phenomena during the evaporation. We discuss important regimes maps reported for different types of deposit, which allow us to classify the deposits and coupled physics. We also present studies that have demonstrated particles sorting in an evaporating bi-dispersed colloidal suspensions on a solid surface. Finally, some remarks for the future research opportunities in this arena are presented.

## 1. Introduction

The evaporation of a droplet containing colloidal particles on a solid surface is a topic of interest in research community from last two decades [1]. Several extensive studies have been performed because of potential applications in inkjet printing [2], surface coating [3], biochemistry analysis [4], biosensor and diagnostics [5], forensic [6] etc. As investigated by Deegan et al. [7], the evaporation mass flux along the liquid-gas interface is non-uniform and is the largest near the contact line (schematically shown in Figure 1(b)), which creates radially outward fluid flow inside the droplet; the contact line of the droplet is pinned due to the microscopic surface roughness. Here, the evaporation take place by diffusion of liquid vapor from a spherical cap like sessile droplet on the surface. Thus, an evaporating droplet dispersed with colloidal particles results in a ring-like particle deposit, due to the advection of the particles by the outward fluid flow [7], as shown in Figure 1(c), showing a zoomed-in SEM image of the ring. The physics of the droplet evaporation is coupled at disparate time and length scales. For instance, the deposition of the droplet involves impact of the droplet at a much faster time scale and followed by a much slower scale for the evaporation. For instance, a nanoliter droplet on a heated fused silica substrate takes around 10 ms for impact and 4 s for the complete evaporation [8] (Figure 2). Table 1 summarizes all possible transport phenomena and Figure 3 shows these processes using a schematic. A general research question to answer is how flow-field in the droplet and other transport phenomena influence the colloidal deposit formed after the evaporation. Various deposit patterns can be observed, as reported in a recent reviews by Larson [9], Sefiane [10] and Parsa et al. [11]. A most common example of deposit pattern is the drying of a droplet with particles, such as a coffee spill, which forms a ring-like pattern on the surface. This is called the "coffee ring" effect, a term coined by Deegan et al. [7]. In the following, we review the recent key studies which investigated role and effect of important problem parameters in the context of the formation of colloidal deposits using evaporation of droplets.

## 2. Role of contact line depinning

Droplet evaporation involves two modes, the constant contact radius (CCR) mode, where the contact radius stays constant with the contact angle decreasing, and the constant contact angle (CCA) mode, where the contact angle remains constant with the contact radius decreasing [12] i.e depinning of the contact line. The CCR mode is more likely to happen on hydrophilic substrates while the CCA mode is more usual on hydrophobic substrates [11]. The combination of the two modes generates various particle deposition patterns, as shown in Figure 4. If the evaporation occurs in CCR mode then a ring-like deposit forms [13] and if it occurs with depinning of contact line i.e. CCA mode then an inner deposit forms [14]. Moreover, if the evaporation occurs with stick-slip motion of contact line (combination of CCR and CCA) then multiple-rings deposit form [15].

Sangani et al. [16] studied the droplet evaporation containing colloidal particles for a wide range of particles diameter ($d_P$ = 1 to 70 μm). They calculated the surface tension force acting on the particle near the contact line and gave a criterion for the contact line pinning. Further, Jung et al. [17] and Wong et al. [18] explained that the deposit patterns depend on the evaporation mechanism of the droplet on the surface. During evaporation, the particles inside the droplet experience several forces, namely, surface tension (or capillary) force, drag force, friction force, and adhesion forces due to the van der Waals force and electrostatic interaction forces between particles and substrate as well as among the particles themselves inside the fluid. The particles deposition near the contact line is the results of the competition between these forces acting on each of the particles near the contact line. During the evaporation as the inward surface tension force overcomes all other forces then depinning of the contact line occurs, otherwise not [16, 19]. There are several studies that shows the contact line pinning and coffee-ring formation [10]. But, from the application point of view, researchers investigated several ways to change or control the coffee ring effect, summarized recently by Parsa et al. [11].

## 3. Effect of Substrate heating

Recently, various research groups investigated that the coffee ring effect can be suppressed by varying surface tension gradients of liquid-gas interface, though droplet was evaporating in CCR mode. It can be caused by varying substrate temperature [20, 21] which induced thermal Marangoni stresses at the liquid-gas interface. This generates a convective flow near the liquid-gas interface from lower surface tension region (droplet contact point) to the higher surface tension region (droplet apex point), called as thermal Marangoni flow. The Marangoni circulation reverses the outward capillary flow of particles and changes the particle deposition from the ring to the droplet center region [22]. Note that a thermal Marangoni flow also exists due to non-uniform cooling of the interface by latent heat of evaporation. Hu and Larson [23] and Bhardwaj et al. [24] reported that such a Marangoni flow can suppress coffee-ring effect.

Recently, few studies were carried out which used substrate heating. For example, Li et al. [20] varied glass ($\theta_{eq}$ = 24 - 30°) surface temperature from $T_s$ = 30 to 80°C and obtained various deposit patterns as shown in Figure 5. They studied evaporation of 2.5 µL water droplets containing 0.25 % (v/v), 100 nm polystyrene particles. The deposit pattern changes from coffee-ring to a "coffee eye" - thin ring with an inner deposit with increase in substrate temperature. In the figure, at heated substrate temperature, the inner deposit increases due to higher Marangoni recirculation inside the droplet, which bring more particles towards droplet center. Further, Parsa et al. [21] varied surface temperature from $T_s$ = 25 to 99°C and obtained uniform deposit pattern to a dual-ring pattern transformation on smooth silicon surface ($\theta_{eq}$ = 30 - 40°) as shown in Figure 6. They used 0.05 % (w/w) copper oxide nanoparticles. They demonstrated that uniform particle deposit changes to dual-ring pattern and the size of the secondary ring changes with increase in substrate temperature. At highest substrate temperature, a stick-slip motion found which give rise to multi-ring patterns. Similarly, Zhong and Duan [25] studied the evaporation of droplets containing graphite nanopowders of $d_P$ = 2-3 nm particles with $c$ = 0.05 % (v/v). The substrate temperature was varied from $T_s$ = 10 to 50°C. They also found uniform deposition of particles changes to dual ring (outer thin ring and inner thick ring) patterns with increase in substrate temperature, shown in Figure 7. Further, Patil et al. [14] studied the effect of substrate heating ($T_s$ = 27 to 90 °C) on hydrophilic glass ($\theta_{eq}$ = 34−38°) and hydrophobic silicon ($\theta_{eq}$ = 94−97°). They considered the evaporation of droplets containing $d_P$ = 0.46 µm polystyrene particles with different particles concentration $c$ = 0.05, 0.1, and 1.0% (v/v), and proposed a regime map showing three deposit patterns, namely, ring, thin ring with inner deposit, and inner deposit, as shown in Figure 8. At ambient, as expected a ring and an inner deposit formed on glass and silicon, respectively. At higher substrate temperature, a thin ring with inner deposit formed on both substrates. In all of above studies, the thinning of the ring and formation of either inner deposit or the dual ring attributed to the Marangoni convection inside the droplet that transports the particles from the contact line region to the inner region of the droplet.

## 4. Effect of particle size

The particle size can also influence the deposit patterns. In this regard, recently various studies were performed on the evaporation of a droplet containing colloidal suspensions of different particle diameters. Perelaer et al. [26] used colloidal suspensions of silica particles of diameter $d_P$ = 0.33, 1, 3 and 5 µm ($c$ = 1% w/w) on different glass substrates ($\theta_{eq}$ = 5-130°). They found that for smaller particle sizes ring-like patterns are formed on lesser contact angle substrates and for larger particle sizes inner deposit or uniform deposit patterns are formed on all tested substrates. Similar type of particle deposit patterns behaviour is recorded by Biswas et al. [27] for polystyrene particles of diameter $d_P$ = 0.02, 0.2 and 1.1 µm ($c$ = 0.05% w/w) on different glass substrates. They showed that the particle size significantly affect the deposit patterns and reported that the particles deposit at the contact line as per its geometrical sizes. Weon and Je [19] studied effect of two particle sizes $d_P$ = 0.1 and 1.0 µm ($c$ = 1% w/w) on glass ($\theta_{eq}$ = 15°). They explained that smaller size particles were easily pinned at the contact line, as they were larger in number and can get closely packed next to each other during evaporation, in comparison to larger size particles which were lesser in number and loosely deposited next to each other at the contact line. In result of that, after complete dry-out smaller size particles form a ring and larger size-particles do not form a ring-like deposit.

Recently, Ryu et al. [28] reported the deposit patterns for particle sizes from 100 nm to 10 µm diameters as shown in Figure 9, and showed that the ring-like deposit is formed for smaller colloids while uniform deposit pattern is formed for larger colloids. Further, they also studied the effect of polyethylene oxide (PEO) polymer concentration in the same solution and even for larger colloids they obtained ring-like deposits. Malla et al. [29] observed the ring-like deposits for all particle sizes ($d_P$ = 0.1 to 3 µm) on hydrophilic glass. However, they reported that in the formed ring the particles morphologies change with the particle size and the particle concentration ($c$ = 0.001 to 1%, v/v). They proposed a regime map of particle deposit morphologies showing discontinuous monolayer ring, continuous monolayer ring, and multiple layers ring.

The effect of particle size coupled with substrate heating has also been reported. Patil et al. [30] investigated the combined effect of substrate temperature ($T_s$ = 27 to 90 °C) and particle size ($d_P$ = 0.1 to 3 µm) on the deposit patterns as shown in Figure 10. They showed that on non-heated silicon substrate ($\theta_{eq}$ = 85°), for smaller size particles inner deposit forms and for larger size particles a ring-like structure forms. While on heated silicon substrate, for all tested particle sizes a thin ring with inner deposit patterns form. They showed that due to combined effect of substrate heating and colloidal suspensions contact line pins on higher contact angle substrate ($\theta_{eq}$ = 85°) and it changes the deposit pattern from an inner deposit to a thin ring with inner deposit.

## 5. Deposition on a Non-wetted surface

Regarding the evaporation of droplets on non-wetted (hydrophobic) surfaces, in general, an inner deposit pattern can be found which is attributed to the CCA mode of evaporation. On such surfaces, the particles stay inside the evaporating droplet with diminishing contact radius and accumulate at the middle or inner range of the droplet. For example, Nguyen et al. [31] obtained the centralized/inner deposit pattern by using a hydrophobic silicon, on which during the evaporation the depinning of contact line occurs bringing the particles to the inner region of droplet, as shown in Figure 11. Before that in another study, Orejon et al. [32] studied the droplet evaporation on non-wetted surfaces and showed that the presence of nanoparticles help in the pinning as well as the stick-slip behaviour of the contact line. Larger concentration of particles can delay the depinning of the contact line in comparison to that of the pure liquid droplet. Further, Li et al. [33] studied a controlled deposition of polymer nanoparticles by changing the contact angle hysteresis (CAH). As shown in Figure 12, the inner deposits formed on weak CAH (such as poly diallyl-dimethyl ammonium chloride or sodium polysulfonate) and ring-like deposits formed on strong CAH (such as polyvinyl pyrrolidone). Recently, Bhardwaj [34] suggested that the internal motion of the particles inside the droplet on a non-wetted surface is from the contact line to the top of the droplet, due to larger evaporation flux on the top of the droplet, as shown by a schematic in Figure 13.

## 6. Formation of a Uniform deposit

For application such as ink-jet printing and manufacturing of bioassays, a uniform deposit is required rather than a ring-like or inner deposit. This has been achieved by varying pH of the suspension in a study by Bhardwaj et al. [35] using a water-glass-titania system. They proposed a regime diagram for three deposit patterns and can be explained by the competition between three characteristic velocities, namely, the radial velocity ($V_{rad}$) caused by highest evaporation rate at the contact line, the Marangoni velocity ($V_{Ma}$) due to variation in the surface tension gradient, and the velocity ($V_{DLVO+}$) due to attractive electrostatic and van der Waals forces between particles and the substrate (called Derjaguin-Landau-Verwey-Overbeek (DLVO) interactions), as shown in Figure 14. A ring was formed when $V_{Ma}$ and $V_{DLVO+}$ is negligible in comparison to $V_{rad}$. A ring with central bump (i.e. inner deposit) was formed when $V_{Ma}$ is comparatively higher than $V_{rad}$ and $V_{DLVO+}$, and uniform deposit was formed when the particles and substrate interactions becomes more dominant than radial and Marangoni velocities.

In another notable report, Still et al. [36] altered the deposition pattern of monodispersed particles (1.33 μm, c = 0.5 w/w %) from the "coffee-ring" to a more uniform pattern by adding a surfactant, sodium dodecyl sulfate (SDS). This was explained by the surfactant-induced Marangoni flow. The surfactant was concentrated near the contact line due to the radial flow and higher evaporation rate closer to the contact line; therefore, the surface tension was decreased locally and induced the Marangoni flow, bringing the particles inward as well as contributing to the depinning of the contact line. Marin et al. [37] investigated surfactant-driven flow inside a sessile droplet during evaporation on a non-heated glass substrate. Two surfactants, polysorbate 80 (P80) and SDS, were studied. Both radial flow and surface tension difference decreased in the presence of P80 and increased significantly in the presence of SDS. Interestingly, Still et al. [36] and Marin et al. [37] suggested the opposite on the correlation of the surfactant-induced Marangoni flow and the final deposition pattern. Still et al. [36] believed the Marangoni flow can lead to a more uniform deposition, while Marin et al. [37] believed there is little correlation due to the "rush-hour effect" at the end stage of the evaporation.

Uniform deposit patterns can also be obtained by varying particle shape [38, 39]. Yunker et al. [38] studied evaporation of droplets loaded with ellipsoidal particles and observed uniform deposit patterns. Because of the anisotropic shape of the particles the interparticle interactions between them generates a loosely packed assembly of particles on the liquid-gas interface and blocks the particles advection to the edge and thus uniform deposit pattern forms. Dugyala and Basavaraj [39] also reported the controlled uniform deposit pattern formation from the ellipsoidal particles which was achieved due to an attractive particle-particle as well as particle-substrate interactions.

## 7. Particles sorting in bi-dispersed suspensions

As discussed above, the deposits of mono-dispersed colloidal particle suspension has been studied vastly, however, in recent years utilizing bi-dispersed colloidal particles has gained attention due to its applications in biochemistry analysis [4], controlled evaporative self-assembly [40] and biosensors [5]. Specifically, few studies focused on to achieve the particles sorting in the deposit. Han et al. [40] observed self-assembly of bi-dispersed particles (50 nm and 500 nm, 1:1, $c$ = 0.01

% w/w), with smaller particles deposited closer to the contact line and larger one adjacent to them, in a cylinder-on-flat geometry. The self-sorting feature was attributed to the gradual decreasing of the liquid-gas interface height towards the contact line. The particle carried by the radial flow towards the contact line stops when its diameter matches the height of the interface. Therefore, smaller particles can be transported closer to the contact line. Similarly, Chhasatia and Sun [41] investigated the particle sorting by considering the effect of substrate wettability. They used ink-based bi-dispersed solution ($d_P$ = 100 nm and 1.1 μm, $c$ = 0.5% v/v each) on coated glass having different receding contact angles ($\theta_{rec}$ = 80, 55, 30, 10, and 0°). They reported with increase in substrate wettability the good amount of sorting was seen on substrates having 0° receding contact angle, as shown in Figure 15. No sorting was seen on substrates having high receding contact angle, as there was formation of inner deposit.

Further, Monteux and Lequeux [42] studied the effect of particle size combination in bi-dispersed suspensions. They used three combinations ($d_P$ = 100 nm and 1 μm, 100 nm and 5 μm, 1 and 5 μm, $c$ = 0.16-1.6% w/w) and observed the particle sorting at the contact line. They observed that at the contact line of the droplet a thin film region exists and depending on the contact angle ($\theta_{eq}$ = 10°) and the particle size. Similar type of thin film region was reported recently by Parsa et al. [43] for bi-dispersed particles (1 μm and 3.2 μm, 1:1, 0.025% w/w) on heated silicon. In this study, they reported the effect of substrate temperature ($T_s$ = 25 to 99°C) during sorting of particles at the contact line. As shown in Figure 16, a thin film exists at the outermost ring, next to that smaller particles deposited at the middle ring and a mixture of smaller and larger particles deposited at the innermost ring. However, they did not study different particle size combinations. Thereafter, Patil et al. [30] investigated the combined effect of particle size combinations and the substrate temperature on self-sorting of bi-dispersed colloidal particles on silicon wafer. They proposed a regime map for the observed deposit patterns, as shown Figure 17. On non-heated silicon, the depinning of the contact line was seen and in result of that mixture of smaller and larger particles was formed. Since heating of a hydrophobic surface induces pinning of the contact line [14], Patil et al. [30] obtained the sorting of particles in such a system for particle diameter ratio smaller than 0.18 and the substrate temperature higher than 60°C. However, in their study only one substrate was studied. Very recently, Iqbal et al. [44] reported bi-dispersed colloidal particles on hydrophilic as well as hydrophobic substrates. To devise the particle sorting mechanism, they performed the force analysis on the bi-dispersed particles near the contact line. Figure 18 shows the schematics of the flow inside the droplet and the obtained final deposit patterns on hydrophilic and hydrophobic substrate. As expected on hydrophilic substrate sorting of 0.2 and 3 µm diameter particles was reported, which was attributed to the dominant radial outward driven flow and the friction forces as compared to the inward surface tension force. While on hydrophobic substrate, a mixture of 0.2 and 3 µm diameter particles was seen, which was attributed to the dominant inward surface tension driven forces in comparison to that of other forces.

## 8. Concluding remarks and future outlook

We have reviewed recent key studies, which investigated the formation of a colloidal deposit via the evaporation of a sessile droplet on a solid surface. Several transport phenomena were found important in controlling the deposit shape. The radial flow due to the maximum evaporation at the contact line, flow of particles towards the substrate due to electrostatics and van der Waals forces, and Marangoni convection due to thermal gradient at the liquid gas interface are primary mechanism to produce a ring-like, uniform and inner deposit respectively. The Marangoni convection could become intense with the substrate heating and coupled with a pinned line, a deposit with a thinner ring and inner deposit form. The shape of the particles has been reported to be important and a droplet with ellipsoid particles result in a uniform deposit rather than a ring. We discussed regimes maps for different deposit patterns reported in the literature. There maps allow us to classify the deposits and understand the underlying mechanism of the formation of the deposit. We have briefly explained the particles sorting achieved by the evaporation of bi-dispersed colloidal suspensions on a solid surface. The sorting depends on the ratio of the diameter of the particles and intensity of the substrate heating.

There exist several opportunities for future research in this context. For instance, this knowledge base will help to explain deposit obtained from the drying of blood droplets which could be useful in forensics [45], biomarkers [46] and fingerprint residues in blood pattern analysis [6]. In biology, the bio-molecular interaction between the particles and substrate is important to understand the formation of bio-assays [47]. In this review, we tried to put recent studies that considered the effect of substrate heating, particle size and substrate wettability, however, the combined or coupled effect of each of these parameters over another for wider range is still lacking in the literature. Further, in the last stage of droplet evaporation, the evaporation mechanism is much more complex and transient since there exists thin film of droplet liquid to evaporate.

However, its evaporation mechanism is little explored and it is important to investigate in future, as this last stage of evaporation (known as "rush-hour" effect) could significantly affect the deposit patterns.

## 9. Acknowledgments

R.B. gratefully acknowledges financial support by a grant (EMR/2016/006326) from Science and Engineering Research Board (SERB), Department of Science and Technology (DST), New Delhi, India.

## 10. References

[1] R. G. Larson, "In Retrospect: Twenty years of drying droplets," *Nature,* vol. 550, p. 466, 2017.
[2] B. J. De Gans, P. C. Duineveld, and U. S. Schubert, "Inkjet printing of polymers: state of the art and future developments," *Advanced materials,* vol. 16, pp. 203-213, 2004.
[3] E. Tekin, P. J. Smith, and U. S. Schubert, "Inkjet printing as a deposition and patterning tool for polymers and inorganic particles," *Soft Matter,* vol. 4, pp. 703-713, 2008.
[4] Y. Cai and B.-m. Zhang Newby, "Marangoni flow-induced self-assembly of hexagonal and stripelike nanoparticle patterns," *Journal of the American Chemical Society,* vol. 130, pp. 6076-6077, 2008.
[5] J. T. Wen, C.-M. Ho, and P. B. Lillehoj, "Coffee ring aptasensor for rapid protein detection," *Langmuir,* vol. 29, pp. 8440-8446, 2013.
[6] S. Shiri, K. F. Martin, and J. C. Bird, "Surface coatings including fingerprint residues can significantly alter the size and shape of bloodstains," *Forensic science international,* vol. 295, pp. 189-198, 2019.
[7] R. D. Deegan, O. Bakajin, T. F. Dupont, G. Huber, S. R. Nagel, and T. A. Witten, "Capillary flow as the cause of ring stains from dried liquid drops," *Nature,* vol. 389, pp. 827-829, 1997.
[8] R. Bhardwaj, J. P. Longtin, and D. Attinger, "Interfacial temperature measurements, high-speed visualization and finite-element simulations of droplet impact and evaporation on a solid surface," *International Journal of Heat and Mass Transfer,* vol. 53, pp. 3733-3744, 2010.
[9] R. G. Larson, "Transport and deposition patterns in drying sessile droplets," *AIChe Journal,* vol. 60, p. 1538, 2014.
[10] K. Sefiane, "Patterns from drying drops," *Advances in colloid and interface science,* vol. 206, pp. 372-381, 2014.

[11]  M. Parsa, S. Harmand, and K. Sefiane, "Mechanisms of pattern formation from dried sessile drops," *Advances in colloid and interface science,* 2018.

[12]  R. G. Picknett and R. Bexon, "The evaporation of sessile or pendant drops in still air," *J. Colloid Interface Sci.,* vol. 61, pp. 336-350, 1977.

[13]  A. P. Sommer, M. Ben-Moshe, and S. Magdassi, "Self-Discriminative Self-Assembly of Nanospheres in Evaporating Drops," *Journal of Physical Chemistry B,* vol. 108, pp. 8-10, 2004.

[14]  N. D. Patil, P. G. Bange, R. Bhardwaj, and A. Sharma, "Effects of Substrate Heating and Wettability on Evaporation Dynamics and Deposition Patterns for a Sessile Water Droplet Containing Colloidal Particles " *Langmuir,* vol. 32, pp. 11958–11972, 2016.

[15]  S. Maheshwari, L. Zhang, Y. Zhu, and H.-C. Chang, "Coupling between precipitation and contact-line dynamics: Multiring stains and stick-slip motion," *Physical review letters,* vol. 100, p. 044503, 2008.

[16]  A. S. Sangani, C. Lu, K. Su, and J. A. Schwarz, "Capillary force on particles near a drop edge resting on a substrate and a criterion for contact line pinning," *Physical Review E,* vol. 80, p. 011603, 2009.

[17]  J.-y. Jung, Y. W. Kim, J. Y. Yoo, J. Koo, and Y. T. Kang, "Forces acting on a single particle in an evaporating sessile droplet on a hydrophilic surface," *Analytical chemistry,* vol. 82, pp. 784-788, 2010.

[18]  T.-S. Wong, T.-H. Chen, X. Shen, and C.-M. Ho, "Nanochromatography driven by the coffee ring effect," *Analytical chemistry,* vol. 83, pp. 1871-1873, 2011.

[19]  B. M. Weon and J. H. Je, "Self-pinning by colloids confined at a contact line," *Physical review letters,* vol. 110, p. 028303, 2013.

[20]  Y. Li, C. Lv, Z. Li, D. Quéré, and Q. Zheng, "From coffee rings to coffee eyes," *Soft matter,* 2015.

[21]  M. Parsa, S. Harmand, K. Sefiane, M. Bigerelle, and R. Deltombe, "Effect of Substrate Temperature on Pattern Formation of Nanoparticles from Volatile Drops," *Langmuir,* vol. 31, pp. 3354-3367, 2015.

[22]  H. Hu and R. G. Larson, "Analysis of the effects of Marangoni stresses on the microflow in an evaporating sessile droplet," *Langmuir,* vol. 21, pp. 3972-3980, 2005.

[23]  H. Hu and R. G. Larson, "Marangoni effect reverses coffee-ring depositions," *The Journal of Physical Chemistry B,* vol. 110, pp. 7090-7094, 2006.

[24]  R. Bhardwaj, X. Fang, and D. Attinger, "Pattern formation during the evaporation of a colloidal nanoliter drop: a numerical and experimental study," *New Journal of Physics,* vol. 11, p. 075020, 2009.

[25]  X. Zhong and F. Duan, "Disk to dual ring deposition transformation in evaporating nanofluid droplets from substrate cooling to heating," *Physical Chemistry Chemical Physics,* vol. 18, pp. 20664-20671, 2016.

[26] J. Perelaer, P. J. Smith, C. E. Hendriks, A. M. van den Berg, and U. S. Schubert, "The preferential deposition of silica micro-particles at the boundary of inkjet printed droplets," *Soft Matter,* vol. 4, pp. 1072-1078, 2008.

[27] S. Biswas, S. Gawande, V. Bromberg, and Y. Sun, "Effects of particle size and substrate surface properties on deposition dynamics of inkjet-printed colloidal drops for printable photovoltaics fabrication," *Journal of Solar Energy Engineering,* vol. 132, p. 021010, 2010.

[28] S.-a. Ryu, J. Y. Kim, S. Y. Kim, and B. M. Weon, "Drying-mediated patterns in colloid-polymer suspensions," *Scientific Reports,* vol. 7, p. 1079, 2017.

[29] L. K. Malla, R. Bhardwaj, and A. Neild, "Analysis of Profile and Morphology of Colloidal Deposits obtained from Evaporating Sessile Droplets," *Colloids and Surfaces A: Physicochemical and Engineering Aspects,* 2019.

[30] N. D. Patil, R. Bhardwaj, and A. Sharma, "Self-Sorting of Bidispersed Colloidal Particles Near Contact Line of an Evaporating Sessile Droplet," *Langmuir,* 2018/05/29 2018.

[31] T. A. Nguyen, M. A. Hampton, and A. V. Nguyen, "Evaporation of nanoparticle droplets on smooth hydrophobic surfaces: the inner coffee ring deposits," *The Journal of Physical Chemistry C,* vol. 117, pp. 4707-4716, 2013.

[32] D. Orejon, K. Sefiane, and M. E. Shanahan, "Stick–slip of evaporating droplets: Substrate hydrophobicity and nanoparticle concentration," *Langmuir,* vol. 27, pp. 12834-12843, 2011.

[33] Y.-F. Li, Y.-J. Sheng, and H.-K. Tsao, "Evaporation stains: suppressing the coffee-ring effect by contact angle hysteresis," *Langmuir,* vol. 29, pp. 7802-7811, 2013.

[34] R. Bhardwaj, "Analysis of an Evaporating Sessile Droplet on a Non-Wetted Surface," *Colloid and Interface Science Communications,* vol. 24, pp. 49-53, 2018.

[35] R. Bhardwaj, X. Fang, P. Somasundaran, and D. Attinger, "Self-assembly of colloidal particles from evaporating droplets: role of DLVO interactions and proposition of a phase diagram," *Langmuir,* vol. 26, pp. 7833-7842, 2010.

[36] T. Still, P. J. Yunker, and A. G. Yodh, "Surfactant-induced Marangoni eddies alter the coffee-rings of evaporating colloidal drops," *Langmuir,* vol. 28, pp. 4984-4988, 2012.

[37] A. Marin, R. Liepelt, M. Rossi, and C. J. Kähler, "Surfactant-driven flow transitions in evaporating droplets," *Soft Matter,* vol. 12, pp. 1593-1600, 2016.

[38] P. J. Yunker, T. Still, M. A. Lohr, and A. Yodh, "Suppression of the coffee-ring effect by shape-dependent capillary interactions," *Nature,* vol. 476, pp. 308-311, 2011.

[39] V. R. Dugyala and M. G. Basavaraj, "Control over coffee-ring formation in evaporating liquid drops containing ellipsoids," *Langmuir,* vol. 30, pp. 8680-8686, 2014.

[40] W. Han, M. Byun, and Z. Lin, "Assembling and positioning latex nanoparticles via controlled evaporative self-assembly," *Journal of Materials Chemistry,* vol. 21, pp. 16968-16972, 2011.

[41] V. H. Chhasatia and Y. Sun, "Interaction of bi-dispersed particles with contact line in an evaporating colloidal drop," *Soft Matter,* vol. 7, pp. 10135-10143, 2011.

[42]  C. c. Monteux and F. o. Lequeux, "Packing and sorting colloids at the contact line of a drying drop," *Langmuir,* vol. 27, pp. 2917-2922, 2011.

[43]  M. Parsa, S. Harmand, K. Sefiane, M. Bigerelle, and R. Deltombe, "Effect of substrate temperature on pattern formation of bidispersed particles from volatile drops," *The Journal of Physical Chemistry B,* vol. 121, pp. 11002-11017, 2017.

[44]  R. Iqbal, B. Majhy, A. Q. Shen, and A. Sen, "Evaporation and morphological patterns of bi-dispersed colloidal droplets on hydrophilic and hydrophobic surfaces," *Soft matter,* vol. 14, pp. 9901-9909, 2018.

[45]  D. Attinger, C. Moore, A. Donaldson, A. Jafari, and H. A. Stone, "Fluid dynamics topics in bloodstain pattern analysis: comparative review and research opportunities," *Forensic science international,* vol. 231, pp. 375-396, 2013.

[46]  D. Brutin, B. Sobac, B. Loquet, and J. Sampol, "Pattern formation in drying drops of blood," *Journal of fluid mechanics,* vol. 667, pp. 85-95, 2011.

[47]  C. Hurth, R. Bhardwaj, S. Andalib, C. Frankiewicz, A. Dobos, D. Attinger*, et al.*, "Biomolecular interactions control the shape of stains from drying droplets of complex fluids," *Chemical Engineering Science,* vol. 137, pp. 398-403, 2015.

# 11. Tables

Table 1: Summary of the possible transport phenomena for an evaporating droplet laden with colloidal particles on a heated surface

| Fluid dynamics | <ul><li>Incompressible and laminar fluid flow</li><li>Viscosity variation due to temperature change</li><li>Laplace forces on the liquid-gas boundary</li><li>Motion of the wetting line during advancing or receding phase</li><li>Marangoni stresses along the liquid-gas interface</li></ul> |
|---|---|
| Heat transfer | <ul><li>Convection inside the drop and conduction in the substrate</li><li>Imperfect thermal contact between the droplet and the substrate</li></ul> |
| Mass transfer | <ul><li>Diffusion of liquid vapor in ambient of the droplet</li><li>Evaporative driven flow inside the droplet</li><li>Cooling of the drop at the liquid-gas interface by latent heat of the evaporation</li></ul> |
| Particles transport | <ul><li>Advection-diffusion of the particles inside the droplet</li><li>Interactions of free surface with growing peripheral deposit of the particles</li><li>Attractive or repulsive forces between the particles and the substrate</li><li>Inter-particle forces</li></ul> |

## 12. Figures

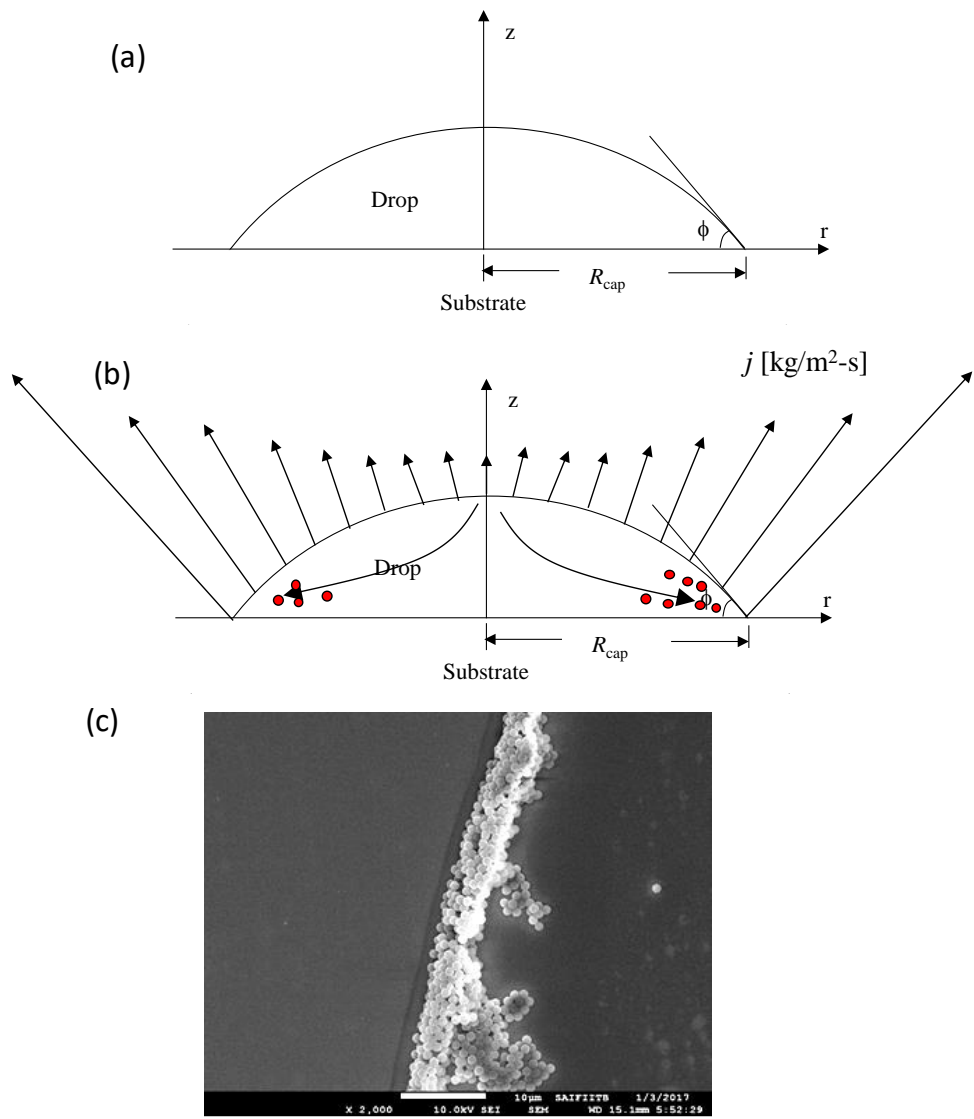

Figure 1: Schematic of an evaporation of a sessile droplet loaded with colloidal particles on a solid surface. The evaporative flux on the liquid-gas interface is shown by vectors, exhibiting the non-uniformity of the flux on the liquid-gas interface. SEM Image in (c) is reprinted with permission from Ref. [30] Copyright (2018) American Chemical Society.

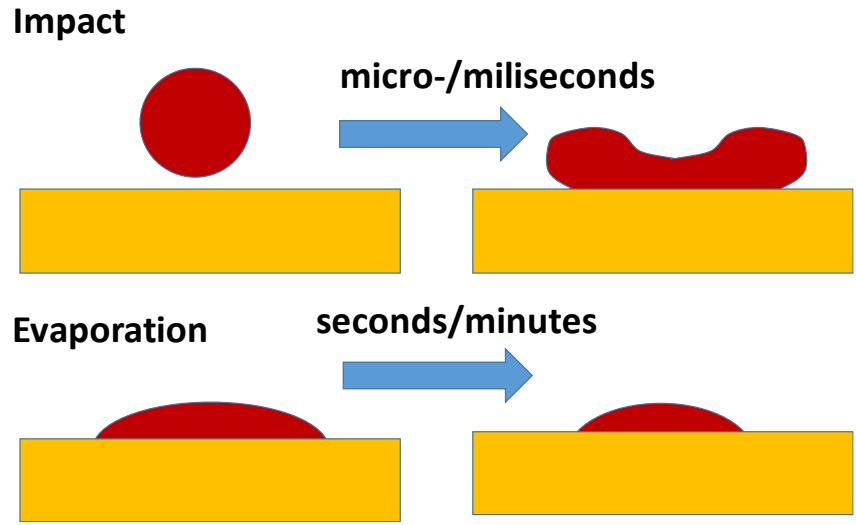

Figure 2: (a) Differences in the time scale of an impacting droplet versus an evaporating microliter or millimeter size water droplet. The evaporation scale is much slower owing to slow diffusion of liquid vapor in the ambient.

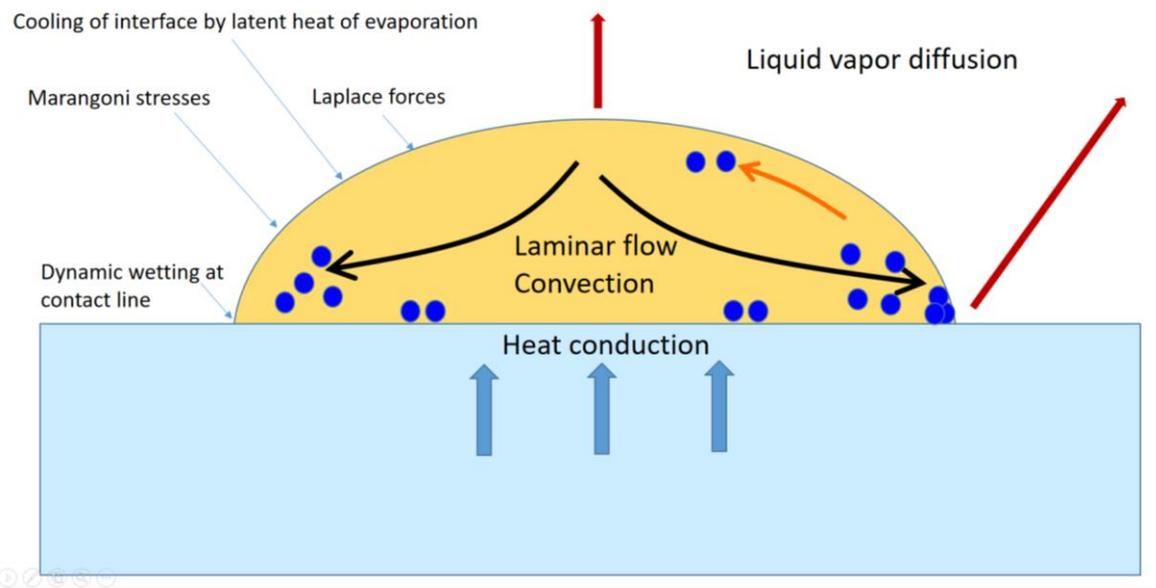

Figure 3: (a) Schematic of an evaporation of a sessile droplet loaded with colloidal particles on a solid surface. Several coupled transport phenomena are shown in the schematic.

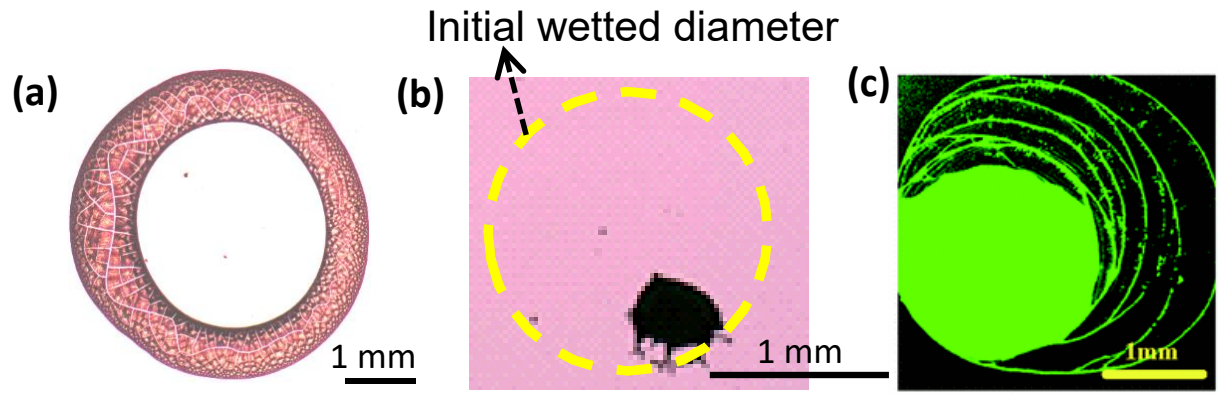

Figure 4: Various particle deposit patterns due to different modes of evaporation- (a) Ring like pattern (Reprinted with permission from Ref. [13] Copyright (2004) American Chemical Society), (b) Inner deposit (Reprinted with permission from Ref. [14] Copyright (2016) American Chemical Society), (c) Multiple ring patterns (Reprinted with permission from Ref. [15] Copyright (2008) American Physical Society).

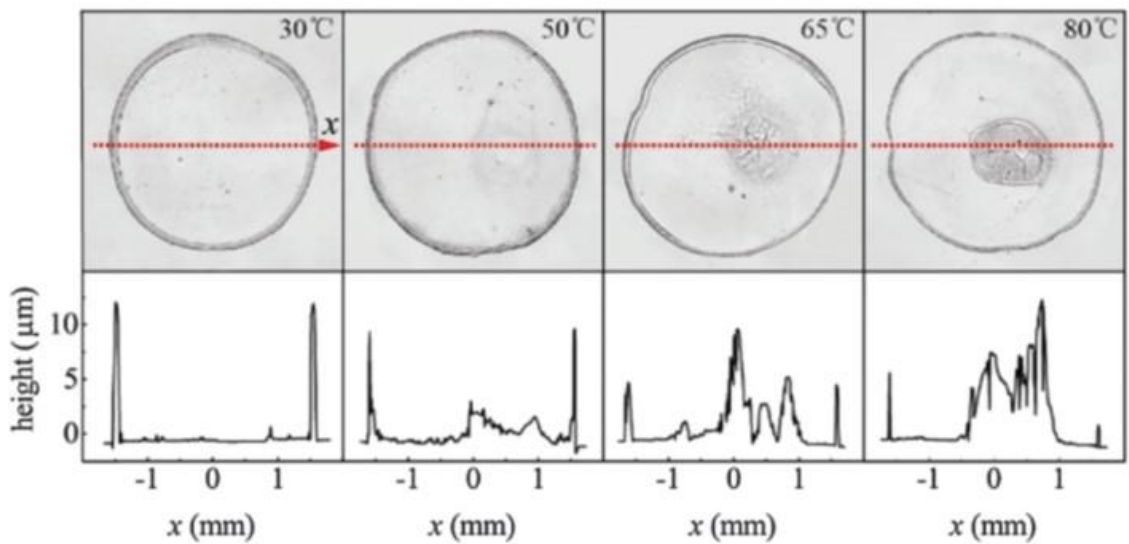

Figure 5: Dried deposit patterns obtained at different glass substrate temperature. Second row shows ring width and height at the corresponding cross-sections measured by interferometry along the red lines. Inner deposit increases with increase in the substrate temperature, and thus the ring becomes thinner. Reprinted with permission from Ref. [20] Copyright (2015) Royal Society Of Chemistry.

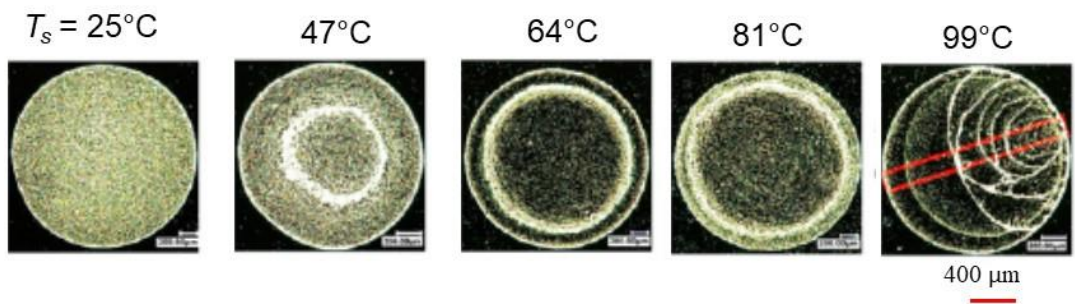

Figure 6: Final deposit patterns obtained on silicon at different surface temperature. Deposit patterns transform from uniform deposits to dual-ring and multi-ring like deposits. Reprinted with permission from Ref. [21] Copyright (2015) American Chemical Society.

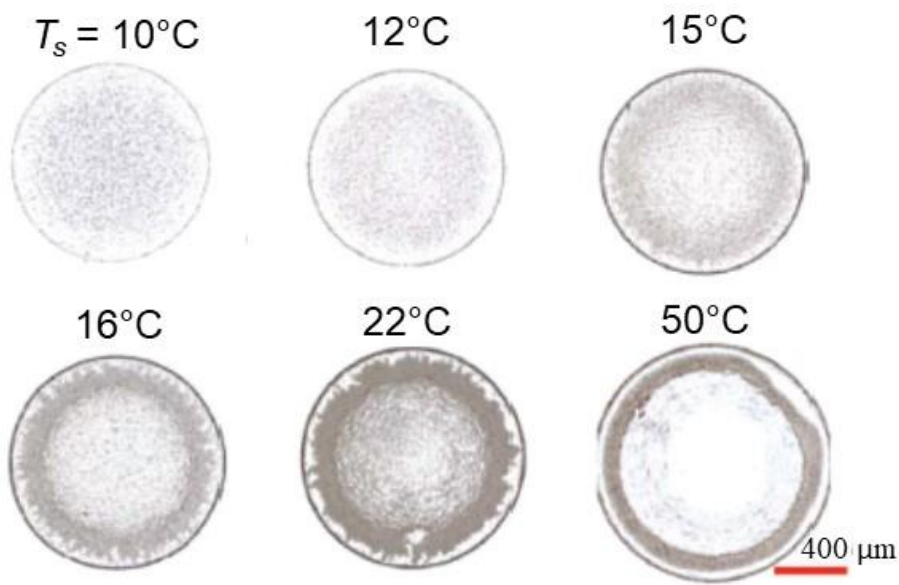

Figure 7: Deposit patterns obtained on silicon wafer as a function of substrate temperature. Deposit patterns transform from uniform deposits to dual-ring deposits. Reprinted with permission from Ref. [25] Copyright (2016) Royal Society Of Chemistry.

Figure 8: Regime map for predicting the deposit patterns as a function of the substrate temperature, substrate wettability and particles concentration. At lower substrate temperature close to ambient, a ring and an inner deposit form on lower and higher contact angle substrates, respectively. At higher substrate temperature, a thin ring with inner deposit forms on all contact angle substrates. Reprinted with permission from Ref. [14] Copyright (2016) American Chemical Society.

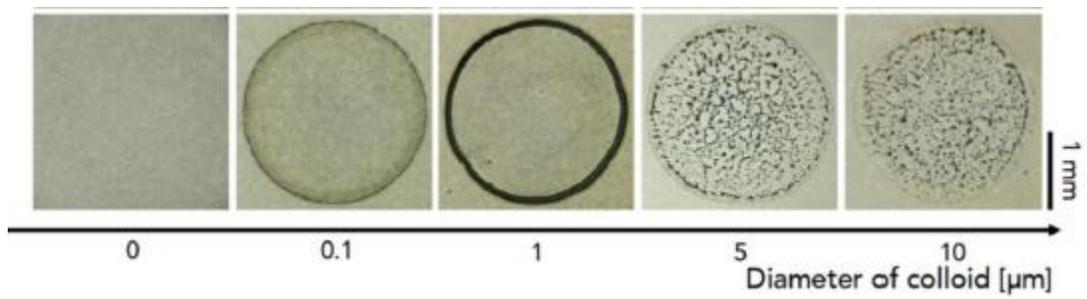

Figure 9: Deposit patterns obtained on the glass substrate showing the influence of the particle size inside the evaporating droplet. Reprinted with permission from Ref. [28] Copyright (2017) Nature Publishing Group.

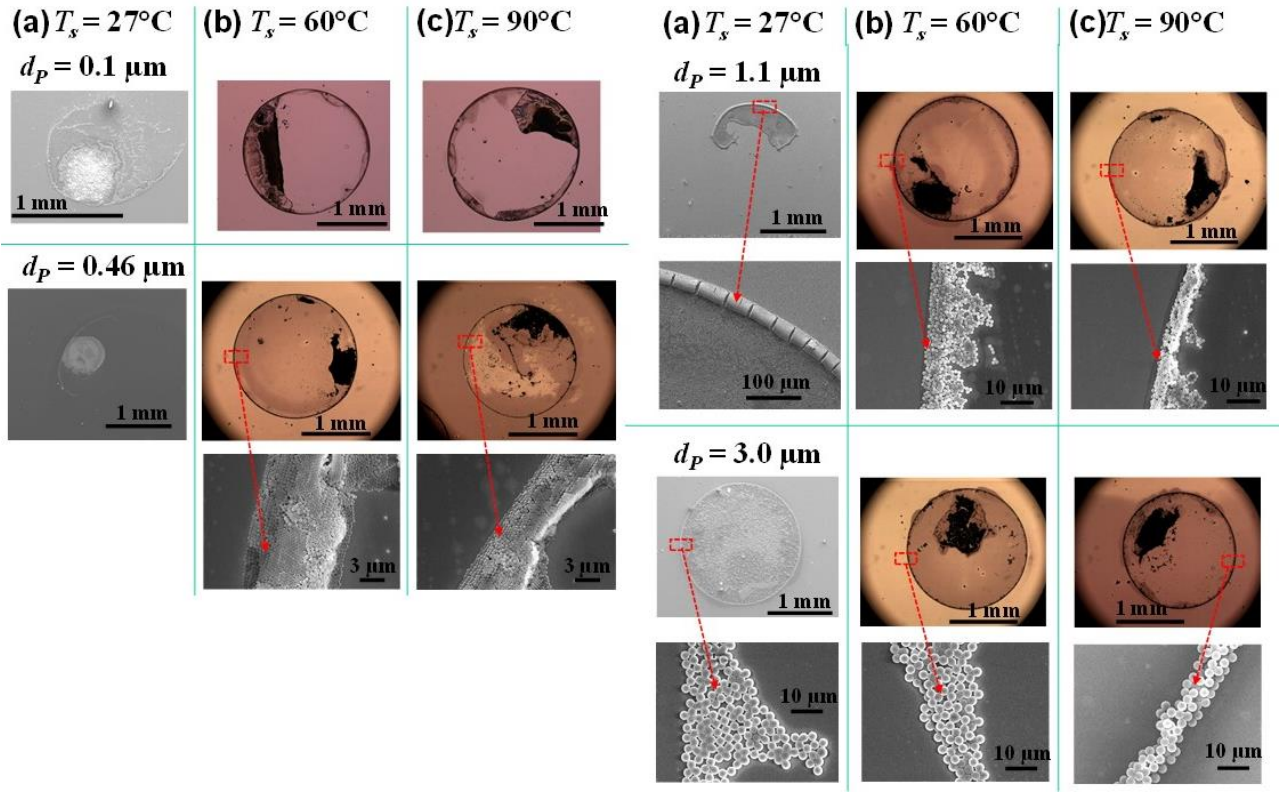

Figure 10: Various particle deposit patterns obtained from the combined effect of substrate temperature and particle size on silicon. On non-heated substrate condition, inner deposit forms for smaller size particles and a ring for larger size particles. On heated substrate condition for all particle sizes, a ring with inner deposit forms. Inner deposit corresponds to the Marangoni convection. Reprinted with permission from Ref. [30] Copyright (2018) American Chemical Society.

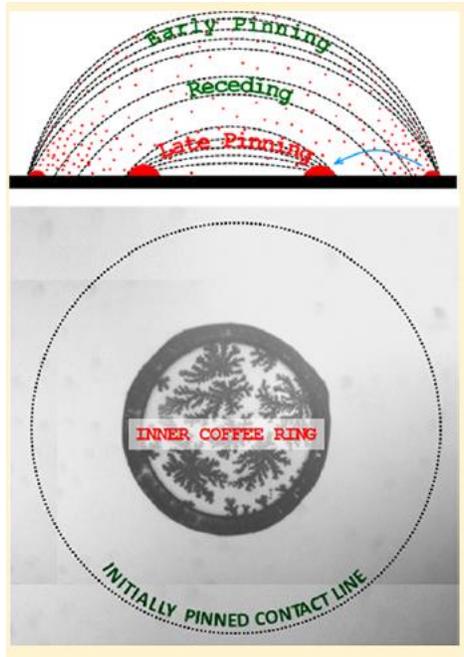

Figure 11: Formation of inner deposit on non-wetted surfaces (hydrophobic silicon) due to the contact line receding. Reprinted with permission from Ref. [31] Copyright (2013) American Chemical Society.

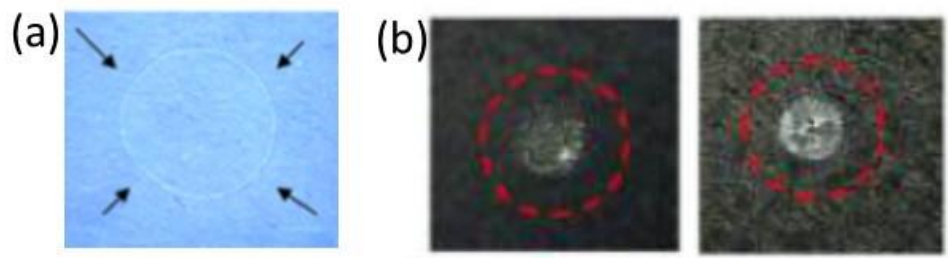

Figure 12: Effect of contact angle hysteresis (CAH) in formation of deposit patterns of polymer particles: (a) a ring on high CAH (b) an inner deposit on weak CAH. Reprinted with permission from Ref. [33] Copyright (2013) American Chemical Society.

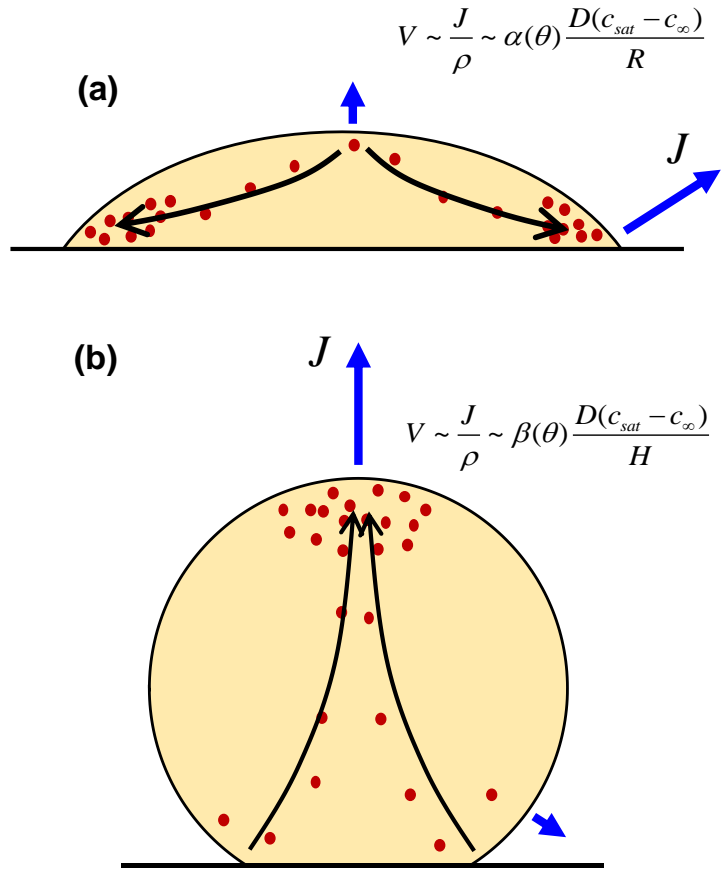

Figure 13: Schematic illustration of the advection of colloidal particles (shown as red dots) by evaporation driven flow on partially wetted (a) and non-wetted (b) surface. Blue arrows compare the magnitude of the evaporation flux near the contact and at the apex of the droplet. The respective velocity scaling is also shown. Reprinted with permission from Ref. [34] Copyright (2018) Elsevier.

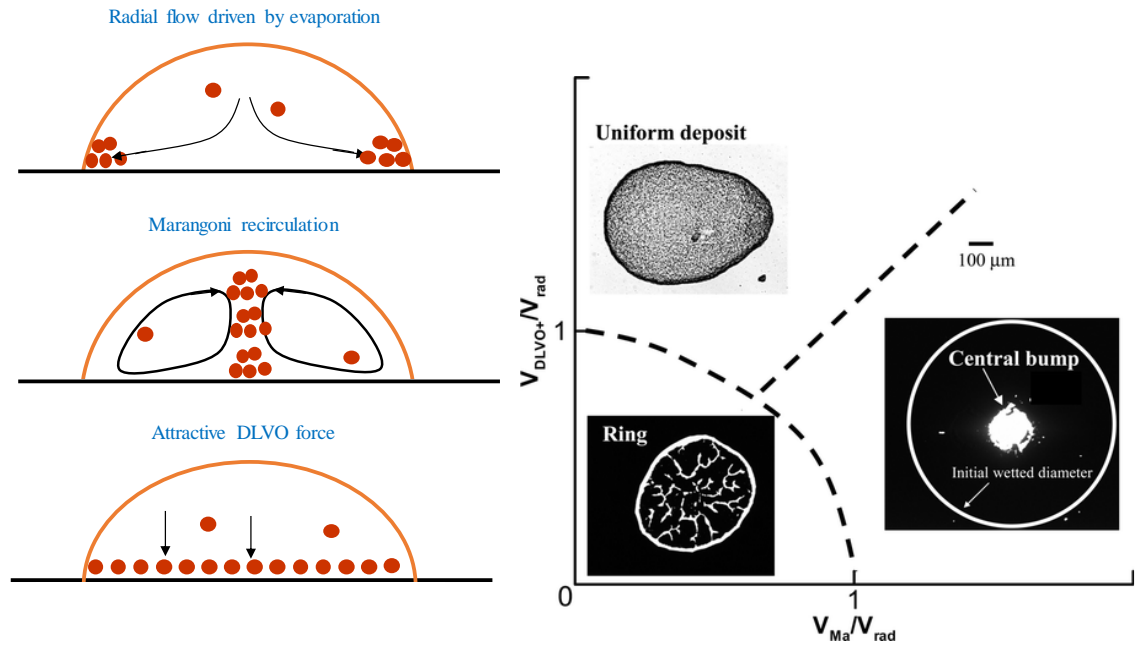

Figure 14: Regime map for the deposit patterns considering the effect of pH of the solution and the evaporation mechanism, which involves the competition between three characteristic velocities ($V_{rad}$, $V_{Ma}$ and $V_{DLVO}$). Reprinted with permission from Ref. [35] Copyright (2010) American Chemical Society.

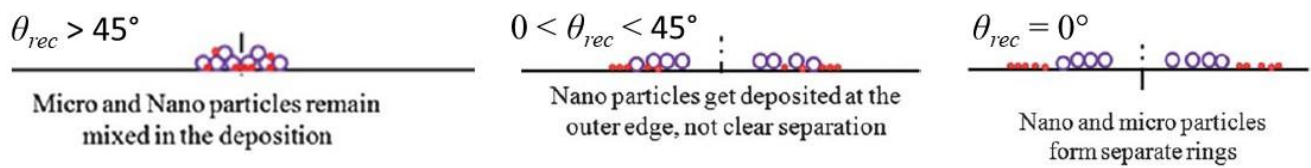

Figure 15: Schematic of the particle sorting onto a glass substrate by systematically varying the substrate wettability/contact angle of substrate. Reprinted with permission from Ref. [41] Copyright (2011) Royal Society of Chemistry.

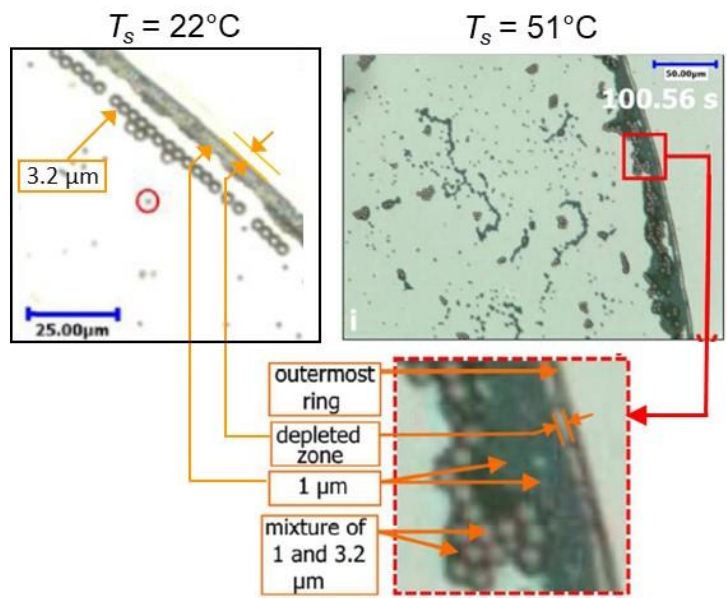

Figure 16: Effect of substrate temperature on particle sorting of bi-dispersed particles from evaporating droplets on silicon. Reprinted with permission from Ref. [43] Copyright (2017) American Chemical Society.

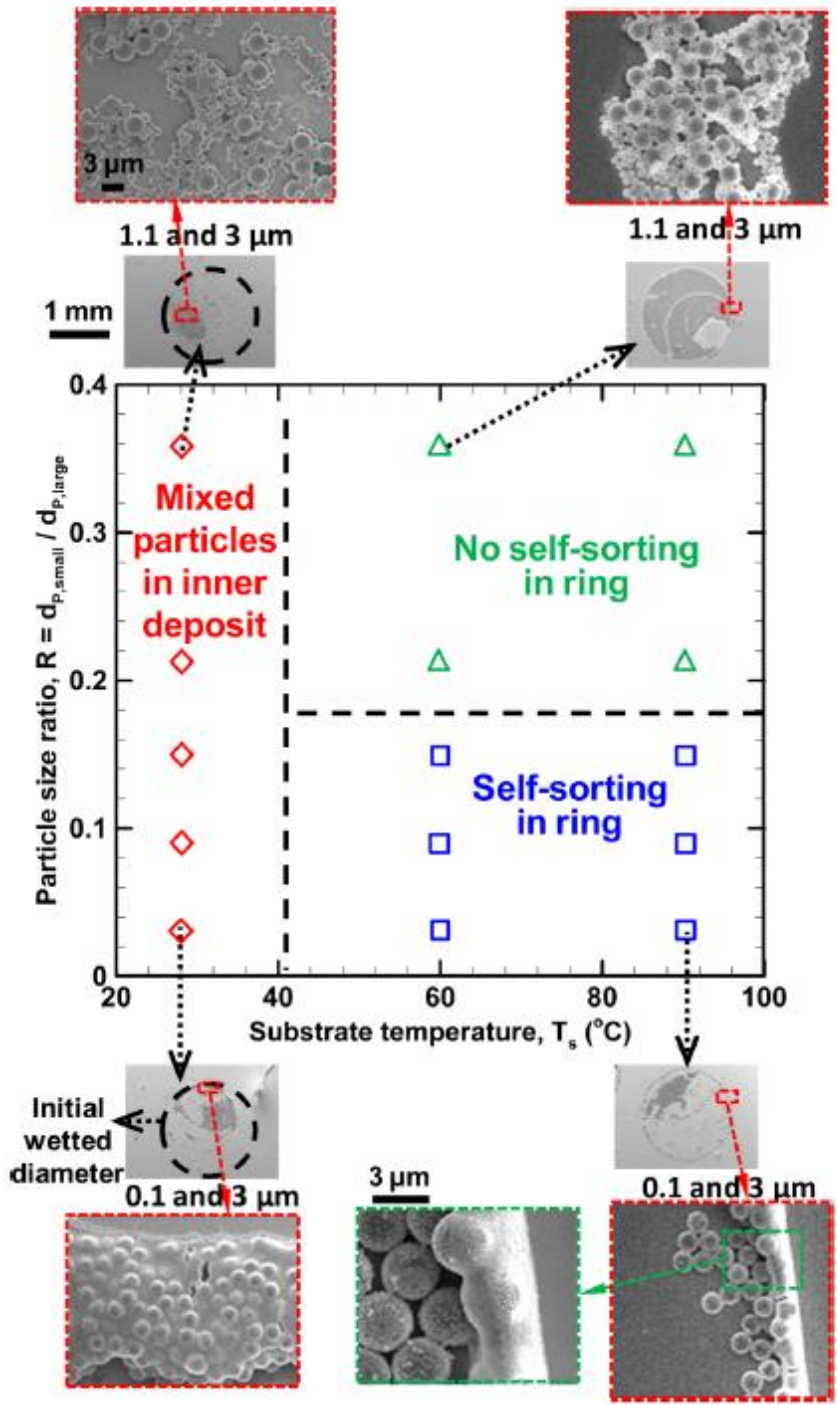

Figure 17: Regime map for the bi-dispersed colloidal particles as a function of the particle size ratio and substrate temperature. Map shows three deposit patterns, namely, mixed particles in inner deposit, self-sorting in ring and no self-sorting in ring. Reprinted with permission from Ref. [30] Copyright (2018) American Chemical Society.

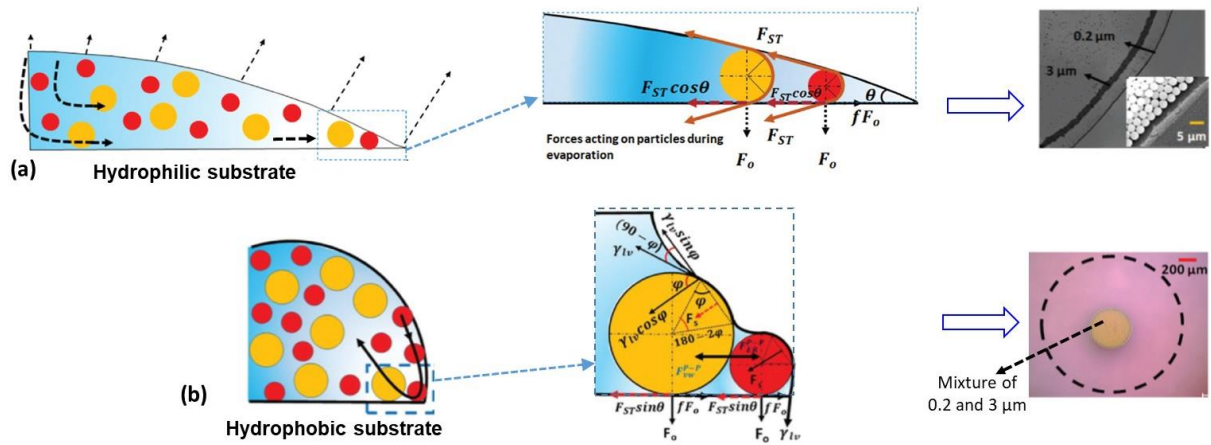

Figure 18: Schematics of the fluid flow inside the evaporating droplet containing bidispersed colloidal particles and the forces acting on the particles near the contact line on (a) hydrophilic and (b) hydrophobic substrates. At the contact line, the particle experiences surface tension force ($F_{ST}$), drag force ($F_D$), friction force ($fF_o$), and adhesion forces ($F_o$) due to the van der Waals force and electrostatic interaction forces between particles and substrate as well as among the particles themselves. (a) Sorting of 0.2 and 3 μm diameter particles was seen, which can be attributed to dominant radial outward driven flow and the friction forces. (b) Mixture of 0.2 and 3 μm diameter particles was seen, which can be attributed to the dominant inward surface tension driven forces. Reprinted with permission from Ref. [44] Copyright (2018) Royal Society of Chemistry.